# Experimental Geometric State Preparation on the Bloch Sphere Via Linear Mechanical Resonator

Abrar N-E Faiaz[1], Donte McMullen [1], Kazi T Mahmood[2], M Arif Hasan[1+*], Pierre Deymier[3+], Keith Runge[3+], Joshua A. Levine[3+]

[1] Department of Mechanical Engineering, Wayne State University, Detroit, MI 48202

[2] Department of Mechanical & Manufacturing Engineering, Bowling Green State University, OH 43403

[3] Department of Materials Science and Engineering, The University of Arizona, Tucson, AZ 85721

[3] Department of Computer Science, The University of Arizona, Tucson, AZ 85721

[+] New Frontiers of Sound Science and Technology Center, The University of Arizona, Tucson, AZ 85721

## Abstract

The Bloch sphere representation, familiar from quantum two-level systems, can also describe classical linear systems whose dynamics are effectively confined to two modes. Here, we experimentally realize this framework in a linear two-mass, three-spring mechanical system by controlling superpositions of its symmetric and antisymmetric eigenmodes. The system is driven from both boundaries at a fixed frequency between the two resonances, so that the steady-state response remains within a two-mode manifold. By independently varying the relative amplitude and phase of the boundary excitations, we generate tunable modal superpositions. Scanning laser vibrometer measurements and frequency-domain analysis are used to extract the complex modal coefficients, which are then normalized and mapped onto Bloch sphere coordinates. This reconstruction reveals planar rotations of the state under phase sweeps and continuous steering between eigenstates under amplitude-ratio tuning. These results provide an experimental validation of Bloch sphere state reconstruction and steady-state geometric state preparation in a minimal linear mass-spring platform, establishing a simple classical testbed for exploring two-mode manifold dynamics.

[*]Corresponding Author: Hasan.Arif@Wayne.Edu

## 1. Introduction

Coupled oscillators provide a canonical setting for studying resonance splitting, mode hybridization, interference, and energy transfer in classical mechanics. More importantly, the equations governing weakly damped coupled modes often reduce to the same mathematical structure as that of driven two-level systems[1]. As a result, classical platforms have long been used to reproduce phenomena more commonly associated with quantum dynamics, including analogues of electromagnetically induced transparency, rapid adiabatic passage, Landau-Zener transitions, and coherent Bloch-type control[1–6]. In recent years, this correspondence has been exploited in mechanical realizations ranging from nanomechanical resonators to macroscopic pendula, demonstrating that several key features of two-level analogue behavior arise from generic wave interference and modal coupling rather than from exclusively quantum ingredients[4,7,8].

A particularly useful representation for describing such systems is the Bloch sphere. When the dynamics of a system are effectively confined to a two-dimensional modal subspace, the complex amplitudes of two orthogonal basis modes can be normalized and interpreted as an effective two-level state. The relative modal weight determines the polar angle, while the phase difference fixes the azimuthal angle, thereby converting an abstract complex superposition into a geometric representation of preparation, rotation, and control[4,8]. In classical settings, this construction is especially valuable because it clarifies which aspects of two-level dynamics follow directly from linear superposition and external driving, and which require genuinely quantum processes such as measurement back-action or spontaneous emission [1,7].

Within acoustics and elasticity, several studies have already shown that mechanical and wave systems can support qubit-like Bloch state representations. Coherent superpositions have been demonstrated in ultrasonic pseudospins, nonlinear granular elastic bits, and more recent elastic-bit architectures that extend Bloch sphere ideas toward Berry-phase and logical-bit constructions[9–12]. These works established that elastic media can support phase-coherent state descriptions and selected gate-like or topological analogies in experimentally accessible settings. At the same time, most of the reported elastic implementations rely on nonlinear response, granular contact mechanics, spectral mixing, or architectures more complex than the minimal two-oscillator problem. However, an important open experimental question remains: can Bloch sphere state reconstruction and steady-state geometric steering be demonstrated directly on the simplest possible linear mechanical platform, where only two normal modes participate, and the state is explicitly tuned by external forcing? Establishing such a platform is valuable for two reasons. First, it isolates the geometric structure of two-mode control from nonlinear frequency conversion, internal beat-driven evolution, and multi-harmonic mode content. Second, it provides a transparent benchmark for the modal-superposition framework that underlies more elaborate elastic-bit and pseudospin systems[9–13].

Here, we address this question experimentally using a linear two-mass, three-spring system driven harmonically from its two outer supports. While previous studies have demonstrated similar concepts using granular systems and non-linear testbeds, these systems often face challenges related to repeatability and stability in experimental front[10,12,14,15]. For example, granular systems lack a restoring force to drive the system back to its initial position, which can lead to difficulties in achieving consistent results and repeatable measurements. In contrast, our system benefits from the restoring forces provided by the springs, which naturally bring the system back to its equilibrium state after each excitation. This ensures that the Bloch sphere trajectories observed in our experiments are consistent and reproducible across multiple trials.

The system is operated at a fixed frequency between the symmetric and antisymmetric resonances so that the steady-state response remains confined to the two-mode manifold spanned by those eigenmodes. By independently controlling the relative amplitude and phase of the two excitations, we prepare a family of modal superpositions, extract their complex coefficients from scanning laser vibrometer measurements, and map the normalized states onto Bloch sphere coordinates. Within this framework, phase sweeps produce planar rotations of the state, whereas amplitude-ratio tuning continuously steers the response between eigenstate-dominated limits. The results provide an experimental realization of Bloch sphere reconstruction and quantum analogous state construction in a minimal linear mass-spring system and establish a simple classical platform for studying two-level state preparation in mechanics.

## 2. Experimental Setup

We designed a mass–spring system consisting of two cylindrical aluminum masses and three springs arranged in a row, as shown in Fig. 1. The two masses (6061 aluminum alloy, 3/4-inch diameter, 1/2-inch long, McMaster-Carr 1610T624) are connected by a zinc-plated music-wire steel spring (1-inch long, McMaster-Carr 9657K309). Two additional springs of the same type are attached to the outer ends of the masses and connected to LDS-210 vibration exciters that drive the system. All springs are fastened to the masses using double-sided tape.

The two exciters are driven by a single B&K Precision 4055B function generator, which provides two synchronized output signals. Using one function generator ensures synchronized excitation at both ends of the lattice and avoids possible desynchronization between separate generators. Each output is sent to a dedicated Brüel & Kjær LDS LPA 100 linear power amplifier before reaching its respective exciter, providing the power required to drive the system. The amplifier settings are matched to preserve symmetry in the excitation.

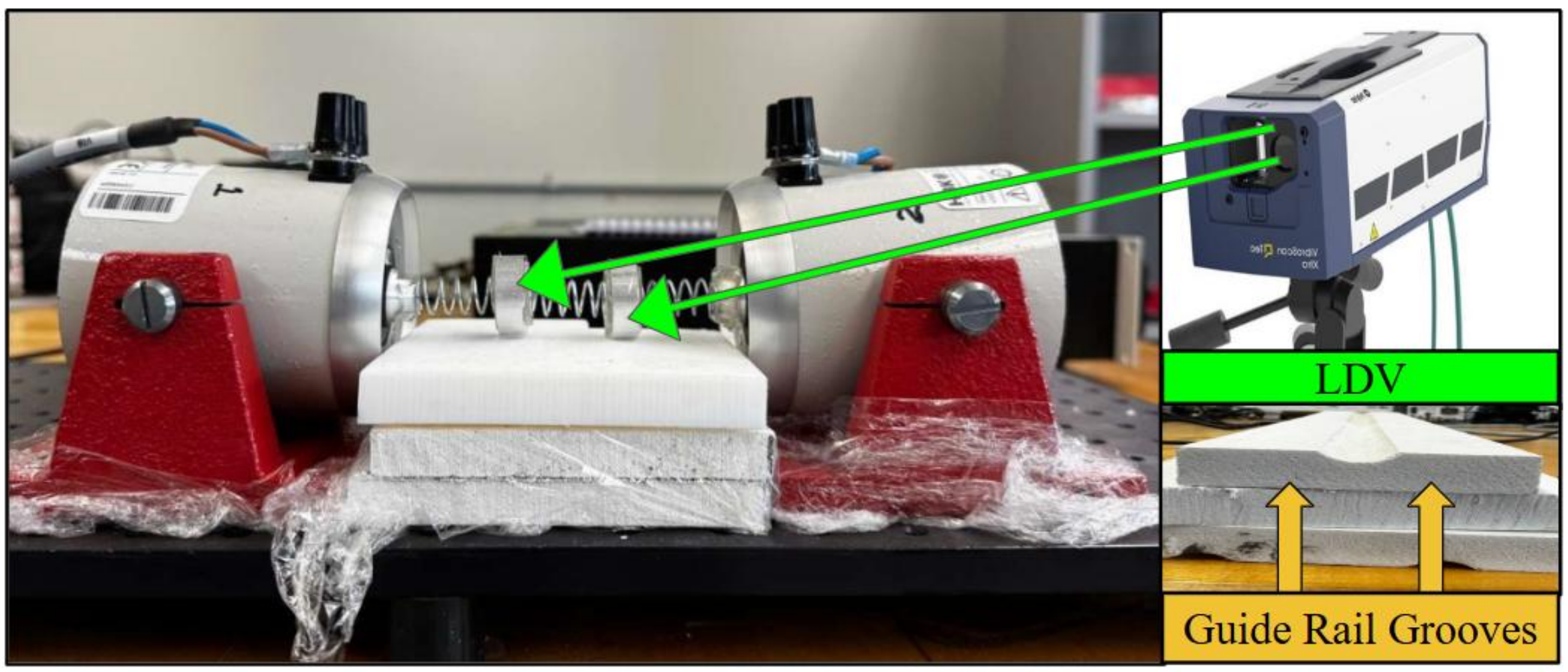

**Figure 1. Experimental Setup for Two-Mass System:** The experimental setup shows the two-mass, three-spring system driven from both boundaries using shakers (left panel). The system is monitored with a Polytec QTec-680 scanning laser-Doppler vibrometer (LDV). The right bottom subplot provides a close-up view of the guide rail grooves used to properly align and position the system for accurate measurements.

To minimize lateral displacement and ensure smooth motion, the assembly rests on a PTFE-based guide rail with a shallow groove that defines the centerline of the motion path (Fig. 1: Right bottom panel). The PTFE surface is large enough to support both masses and is coated with a thin layer of machine oil to reduce sliding friction during vibration. To further isolate the setup from environmental disturbances, the exciters are bolted to an aluminum breadboard, which is secured to a wooden tabletop with vice grips, while rubber dampers are placed beneath the breadboard supports to attenuate external vibrations.

To characterize the intrinsic dynamical properties of the coupled two-mass system, we performed transmission measurements by harmonically driving the right boundary and sweeping the excitation frequency from 20 to 160 Hz. The function generator supplied an input signal of 2 $V_{pp}$, while each amplifier was set to 0.7, corresponding to an effective tenfold amplification of the input signal delivered to the exciters. This particular value of input signal was chosen because increasing the input beyond this level could push the system into the nonlinear regime, where higher-order effects would become more pronounced and distort the measurements.

At each driving frequency, the motion of the two cylindrical masses was measured using a Polytec QTec-680 scanning laser-Doppler vibrometer (SLDV). The laser beam was oriented at an oblique angle of approximately 15° to access the measurement points, and the recorded velocities were corrected for the projection factor during post-processing to recover the longitudinal component of motion. Data was sampled at 200 kHz, with each run lasting 30 s. Prior to data acquisition, a brief stabilization period (10 s) was allowed for the system to settle and reach steady-state conditions after applying the boundary excitation. Following this stabilization period, 10 s of velocity data were recorded for each mass. This acquisition procedure ensures that the system response was fully stabilized and representative of the steady-state harmonic behavior.

A short interval was allotted for repositioning the laser between measurement points. Although the LDV measurements for the two masses were acquired sequentially due to laser repositioning, phase consistency was maintained by recording the synchronized driver reference signal during each acquisition. The phase of each mass response was extracted relative to this common reference through FFT analysis, and the inter-mass phase difference was subsequently reconstructed from these referenced phases. Because both exciters were driven by phase-locked outputs from the same function generator, no independent phase drift existed between acquisitions.

During the experiment, we took several steps to ensure the accuracy and repeatability of our measurements. The LDV setup was carefully calibrated in "2D Alignment" feature to ensure precise velocity measurements, and any slight displacement of the setup required re-alignment to maintain calibration accuracy. Additionally, we maintained a regular maintenance schedule for the equipment, including changing the machine oil used to reduce friction on a weekly basis, ensuring consistent performance.

It is well known that amplifiers, pre-amplifiers and signal generators exhibit so called "harmonic distortion", their amplitude was significantly smaller than the primary driving amplitude. However, these higher harmonics are still present and may contribute to small deviations, though their impact on the overall linearity of the system is minimal. Additionally, AC current equipment inherently operates at a frequency of 60 Hz, which we deliberately avoided during testing to minimize potential interference from this external frequency.

## 3. Experimental Results

### 3.1 Transmission Spectrum and Eigenmode Identification

To characterize the intrinsic dynamics of the coupled two-mass system, we performed transmission measurements by harmonically exciting the boundary and sweeping the driving frequency from 20 to 160 Hz. At each frequency, the system was allowed to reach steady state, and the velocity amplitudes of both masses were recorded and analyzed. The measured amplitude and phase responses are shown in Fig. 2.

The top panel presents the velocity amplitudes of Mass 1 and Mass 2 as functions of frequency. Two pronounced resonance peaks appear near 65 Hz and 130 Hz, which correspond to the two natural modes of the coupled three-spring, two-mass system. These two resonances define the fundamental modal structure of the lattice.

The middle panel shows the phase difference between the two masses. Around the first resonance, near 65 Hz, the phase difference remains close to zero, indicating that the two masses move nearly together. This identifies the first mode as the in-phase, or symmetric, mode, with eigenvector $E_1 = \frac{1}{\sqrt{2}}\begin{pmatrix}1\\1\end{pmatrix}$. Around the second resonance, near 130 Hz, the phase difference increases toward $\pi$, showing that the masses move in opposite directions. This corresponds to the out-of-phase, or antisymmetric, mode, with eigenvector $E_2 = \frac{1}{\sqrt{2}}\begin{pmatrix}1\\-1\end{pmatrix}$.

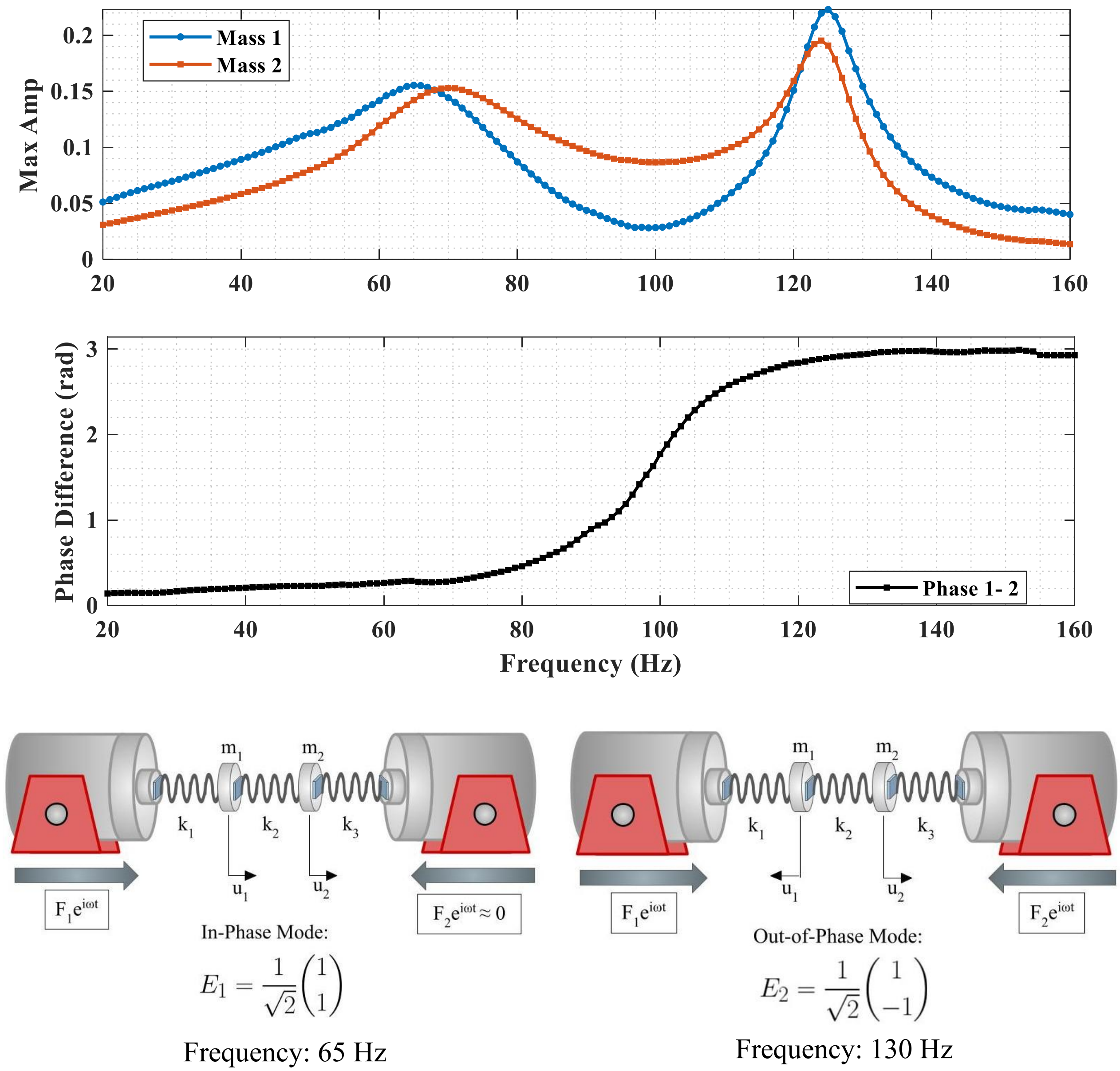

**Figure 2: Transmission characterization of the two-mass three-spring system.** Top: Steady-state amplitude response of both masses under harmonic excitation (2 $V_{pp}$) applied at one boundary, showing two resonant peaks at ~65 Hz and ~130 Hz. Middle: Phase difference between the masses, identifying the in-phase (0) and out-of-phase ($\pi$) normal modes. Bottom: Schematic representation of the corresponding symmetric and antisymmetric eigenvectors.

In an ideal undamped linear system, the phase relation would switch sharply between these two limiting values, with the modal response exhibiting an almost abrupt transition from 0 to $\pi$ across the resonance region. In the experiment, however, this transition occurs gradually over frequencies rather than sudden change. This smoothing is primarily governed by damping in the system. To reduce stick–slip motion and ensure smoother sliding of the masses along the PTFE guide rail, a thin layer of machine oil was applied. While this lubrication improves the quality of motion, it also introduces additional dissipation, effectively increasing the damping of the system. Physically, damping broadens the resonance and spreads the phase evolution over a wider frequency range, so the shift from in-phase motion to out-of-phase motion appears as a continuous rotation rather than an instantaneous jump. Higher damping produces a more gradual and smooth phase transition, whereas lower damping leads to a steeper and more sudden change, approaching the ideal sharp jump. For this reason, the measured phase difference does not remain exactly at 0 near the first mode or exactly at $\pi$ near the second mode but instead approaches these values over a finite frequency interval.

To verify the structural symmetry of the system, the transmission measurement was repeated with excitation applied from the opposite boundary under the same conditions. The resulting resonance frequencies and phase relationships were nearly unchanged, confirming that the mechanical configuration is effectively symmetric and that the identified two-mode structure is robust. Together, these measurements establish a well-defined two-mode basis that serves as the foundation for the subsequent Bloch sphere representation of the system states.

### 3.2 Transition to Steady-State Geometric State Preparation

Having established the symmetric ($E_1$) and antisymmetric ($E_2$) eigenmodes of the system, we next turn to steady state preparation of coefficient of superposition of states. Our objective is to steer the system within the two-mode subspace through boundary driving and thereby realize tunable construction of the elastic state. To enable coherent superposition of the two eigenmodes, we select a driving frequency of 99 Hz, located between the two resonances. At this frequency, neither $E_1$ nor $E_2$ dominates; instead, the steady-state response consists of a mixed modal state, providing access to the full two-dimensional modal manifold. This intermediate regime serves as the operating point for state construction.

To explore steady-state geometric state preparation, we introduce two independent control parameters: the state-ratio parameter $\epsilon$ and the relative driving phase $\delta$. These parameters determine the boundary forcing conditions applied at the two ends of the system and thereby dictate the trajectory of the elastic state within the modal subspace. The detailed theoretical derivation showing how the chosen forcing condition produces the reported Bloch trajectories is provided in the supplementary material.

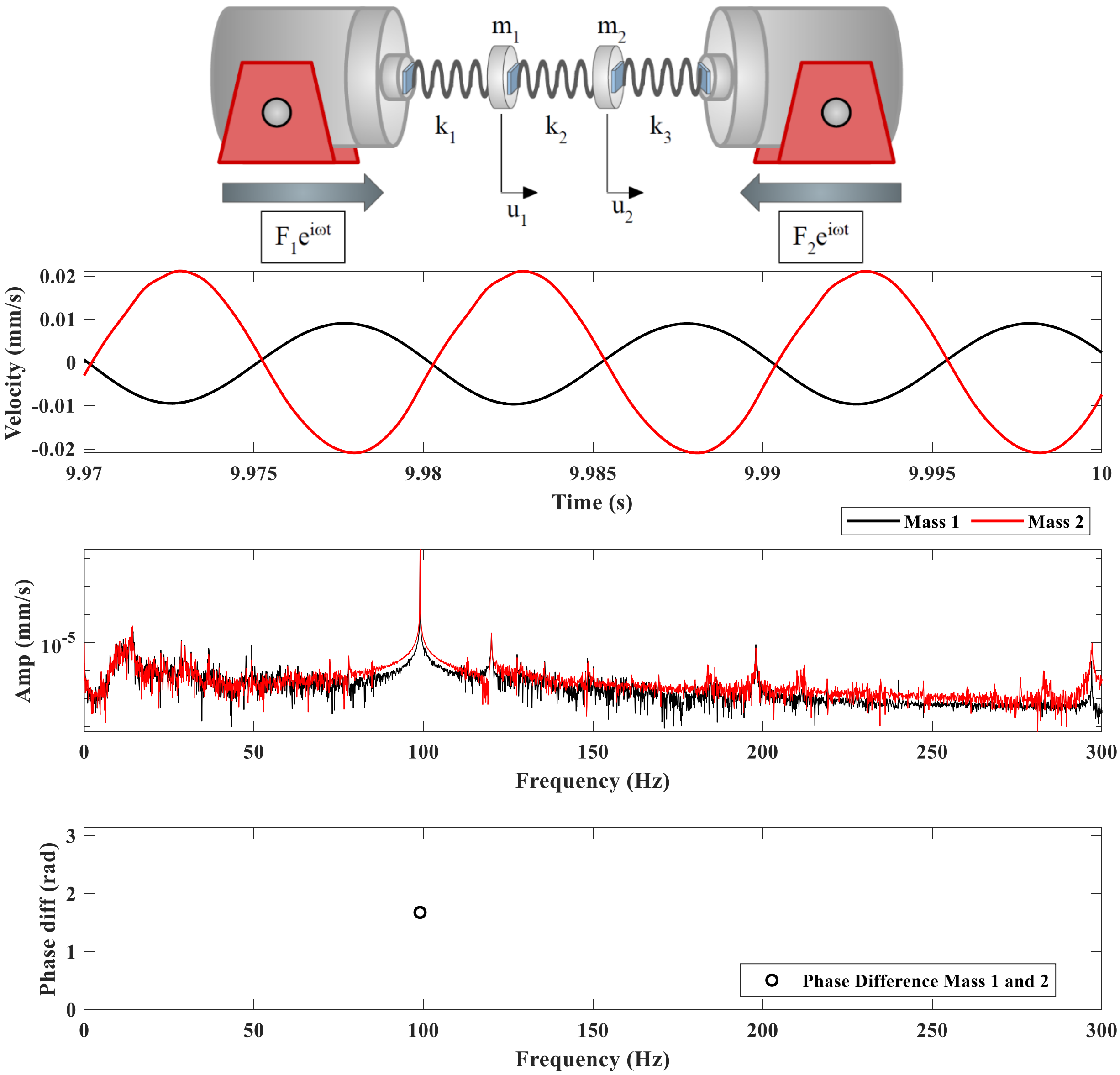

**Figure 3. Time-series and frequency-domain analysis of the two-mass system driven at 99 Hz:** Top: Time-series data for Mass 1 and Mass 2 showing their Velocity oscillations over a 30 $ms$ window. The two masses exhibit phase difference of approximately $\pi/2$. Middle: FFT amplitude spectra for both masses, with a dominant peak at the driving frequency (99 Hz) and significantly smaller peaks for higher-order harmonics, confirming the system's linear response. Bottom: Phase difference between Mass 1 and Mass 2, showing a phase difference of approximately $\pi/2$ radians at 99 Hz, which corroborates the response of velocity oscillation phase difference.

The complex driving conditions applied at the two boundaries are defined by the unscaled target drive vector

$$\tilde{\mathbf{F}}(\epsilon,\delta) = \begin{pmatrix}\tilde{F}_1 & \tilde{F}_2\end{pmatrix} = (\epsilon + (1-\epsilon)e^{i\delta} \quad \epsilon - (1-\epsilon)e^{i\delta}) \tag{1}$$

In this formulation, $\tilde{F}_1$ denotes the complex target drive applied at the right boundary, while $\tilde{F}_2$ denotes the complex target drive applied at the left boundary. The parameter $\epsilon$ controls the relative weighting of the two excitation components, while $\delta$ sets their relative phase. Together, $\epsilon$ and $\delta$ determine the relative amplitudes and phases of the two boundary excitations used to steer the steady-state response within the two-mode manifold.

In the experimental implementation, the quantities in Eq. (1) were used to determine the relative amplitude and phase settings of the two synchronized function-generator channels and not normalized to impose a fixed Euclidean forcing norm. The experimental protocol fixed the maximum channel voltage rather than the total squared drive amplitude. Specifically, the voltage amplitude and phase assigned to channel $j$ were chosen as

$$V_j(\epsilon,\delta) = V_{\max}\frac{\left|\tilde{F}_j(\epsilon,\delta)\right|}{\max\left(\left|\tilde{F}_1(\epsilon,\delta)\right|,\left|\tilde{F}_2(\epsilon,\delta)\right|\right)}, \qquad \gamma_j(\epsilon,\delta) = \arg\left[\tilde{F}_j(\epsilon,\delta)\right], \tag{2}$$

where $V_{\max}$ is the maximum allowed function-generator voltage for the measurement. Thus, the largest applied channel voltage was fixed at $V_{\max}$, while the other channel was scaled proportionally so that the relative complex drive specified by Eq. (1) was preserved. This maximum-channel-amplitude protocol is analogous to driving a multi-mass eigenvector by scaling all channels relative to the largest component.

Because no Euclidean normalization factor was applied, the unscaled drive vector in Eq. (1), $\left|\tilde{F}_1\right|^2 + \left|\tilde{F}_2\right|^2 = 2[\epsilon^2 + (1-\epsilon)^2]$ depends on $\epsilon$. Therefore, the present protocol fixes the maximum commanded channel amplitude and not the total drive norm or injected mechanical energy.

The normalization used later in the Bloch-sphere reconstruction is therefore a state normalization, not a force normalization. After the steady-state response is measured, the extracted modal coefficients are normalized to obtain the Bloch-sphere state vector. This removes the overall response scale and retains the relative modal amplitude and phase that determine the polar and azimuthal Bloch angles.

The system was driven at a frequency of 99 Hz using sinusoidal inputs whose relative amplitudes and phases were determined by $\epsilon = 0.5$ and $\delta = 0$. The complex target values from Eq. (1) were converted into function-generator voltage amplitudes and phases using the maximum-channel-amplitude protocol in Eq. (2).

As shown in the top panel of Fig. 3, the time-series data for Mass 1 and Mass 2 reveal oscillations with a phase difference close to $\pi/2$, which is typical for systems driven at frequencies between their in-phase and out-of-phase eigenfrequencies. Specifically, the system's in-phase eigenfrequency is approximately 65 Hz, and the out-of-phase eigenfrequency is around 130 Hz. The driving frequency of 99 Hz is near the midpoint between these eigenfrequencies, resulting in a phase difference of approximately $\pi/2$.

In the second panel of Fig. 3, the FFT amplitude spectra for both masses shows a dominant peak at 99 Hz, indicating the driving frequency. The amplitude at the driving frequency is significantly larger compared to the higher harmonics, and the closest harmonic peak is approximately $1000$ times smaller in amplitude. This suggests that the system is almost linear, with only a small contribution from higher-order harmonics. The sharp decay of higher harmonics confirms that the system exhibits linear dynamics with minimal distortion.

Another measure taken to ensure the linear dynamics of the system is the frequency sweep performed with a 2 $V_{pp}$ output from the signal generator, applied at one boundary. This ensured a stable phase response and a linear progression of amplitudes, supporting the claim that the system operates linearly at this excitation level, as shown in Fig. 2. For the amplitude sweep, we carefully controlled the input to keep the system within the linear regime. Specifically, we limited the maximum output from each boundary excitation to 1 $V_{pp}$. When both boundaries deliver equal excitation, each channel remains within the selected maximum voltage limit. This ensured that all measurements used for the Bloch-sphere reconstruction were performed within the experimentally verified linear operating regime. Additional quantitative demonstrations of linearity across the full range of forcing conditions used for the Bloch-sphere experiments, including amplitude and phase responses, are provided in the Supplementary Material under the subsection "Linearity of the Two-Mass System".

The phase difference between Mass 1 and Mass 2, as shown in the third panel of Fig. 3, indicates that at 99 Hz, the phase difference for the primary frequency peak ($\omega_D$= 99 Hz) is nearly $\pi/2$ radians. The time-series plot further corroborates this observation, where the oscillations of Mass 1 and Mass 2 exhibit a phase difference of nearly $\pi/2$ radians, consistent with the primary frequency peak at $\omega_D = 99$ Hz. This confirms that the system is predominantly linear at this driving frequency, with minimal contributions from harmonic frequencies.

The system's velocity data, which has been recorded as a function of time, can be expressed in terms of a superposition of two eigenmodes, $E_1$ and $E_2$, corresponding to the in-phase and out-of-phase modes of the system. These modes are represented as:

$$\mathbf{V}(\mathbf{t}) = \begin{pmatrix} v_1 \\ v_2 \end{pmatrix} = \begin{pmatrix} A_1 \\ A_2 \end{pmatrix} e^{i\omega_D t} \equiv (\alpha|E_1\rangle + \beta|E_2\rangle)e^{i\omega_D t} \tag{3}$$

where $\alpha$ and $\beta$ are the complex modal coefficients associated with the eigenmodes $|E_1\rangle$ and $|E_2\rangle$, respectively, and $\omega_D$ is the driving frequency. $A_1 =| A_1 | e^{i\sigma_1}$ and $A_2 =| A_2 | e^{i\sigma_2}$ denote the complex steady-state response amplitudes of masses 1 and 2. From the FFT analysis and phase analysis (shown in Fig. 2), we extract the response coordinate of physical masses $|\mathrm{A}_1|$ and $|\mathrm{A}_2|$ and absolute phases $\sigma_1$ and $\sigma_2$ of each respective mass. These are used to calculate the modal coefficients $\alpha$ and $\beta$ as: $\alpha = \frac{A_1+A_2}{\sqrt{2}}, \beta = \frac{A_1-A_2}{\sqrt{2}}$. Because the steady state response of the system is confined to the two-mode subspace spanned by $|E_1\rangle$ and $|E_2\rangle$, the velocity field can be completely described by the complex modal coefficients $\alpha$ and $\beta$, which encode the relative amplitude and phase of the two eigenmodes.

In Eq. (3), Dirac notation is adopted as a compact representation for vectors in the system's state space[11]. Although commonly used in quantum mechanics, this notation provides a convenient way to describe the elastic states of the present system because the modal vectors $|E_1\rangle$ and $|E_2\rangle$ are mutually orthogonal eigenmodes that form an orthonormal basis for a two-dimensional state space. Within this basis, the velocity of the system can be expressed as a superposition of these two modes.

The complex coefficients $\alpha$ and $\beta$ encode both the relative amplitude and phase of the modal contributions and therefore describe a coherent superposition between the two possible vibration patterns $|E_1\rangle$ and $|E_2\rangle$. Consequently, the modal contribution to the total velocity field can be represented by a column state vector describing the velocity state of the system.

This framework enables the creation of a two-level mechanical subsystem, referred to as an *elastic bit*, through externally applied drivers. An elastic bit serves as a classical analogue of the superposition property of a qubit, the fundamental information unit in quantum computing platforms. In this work, we represent the states of the elastic bit geometrically using the Bloch sphere. Each elastic bit represents a vector originating from the center of the sphere and pointing to the surface, parameterized by two angles: the polar angle $\theta$ and the azimuthal angle $\phi$. These angles encode the relative modal amplitude and phase, allowing the Bloch sphere to represent the superposition of the two eigenstates $|E_1\rangle$ and $|E_2\rangle$[12]. The detailed derivation of this mapping, along with the equations used to obtain the complex amplitudes $A_1$ and $A_2$, is provided in the supplementary material. This mathematical framework allows the system's state to be represented geometrically on the Bloch sphere, where the angles $\theta$ and $\phi$ fully describe the state of the system. The evolution of the system can then be interpreted as a trajectory on the Bloch sphere, where the state vector moves, with its head tracing an arc on the sphere's surface. The orientation of this vector specifies the elastic state of the system.

Accordingly, the Bloch sphere parameters used in this study are obtained from the modal coefficients as:

$$\theta = 2\cos^{-1}(|\ \alpha\ |), \phi = \arg(\alpha) - \arg(\beta)$$

This representation allows the elastic state of the system to be visualized geometrically on the Bloch sphere, where the angles $\theta$ and $\phi$ specify the orientation of the state vector. The state is therefore represented by an arrow originating from the center of the sphere (shown as the curved arrow on the equatorial plane in the left panel of Fig. 4). The evolution of the system can then be interpreted as a trajectory on the Bloch sphere, obtained by varying the control parameters $\epsilon$ and $\delta$, which determine the relative amplitude and phase of the boundary driving forces.

In the first scenario, we kept the state-ratio parameter $\epsilon$ constant and varied the relative driving phase $\delta$ from $-\pi$ to $\pi$. This configuration is illustrated in the left panel of Fig. 4. As $\delta$ varies, the system state traces a complete revolution within a single plane on the Bloch sphere, starting and ending at the same point. This behavior reflects the system's oscillatory nature, where the relative phase between the two driving forces controls the trajectory of the system's state vector, forming a circular path on the Bloch sphere.

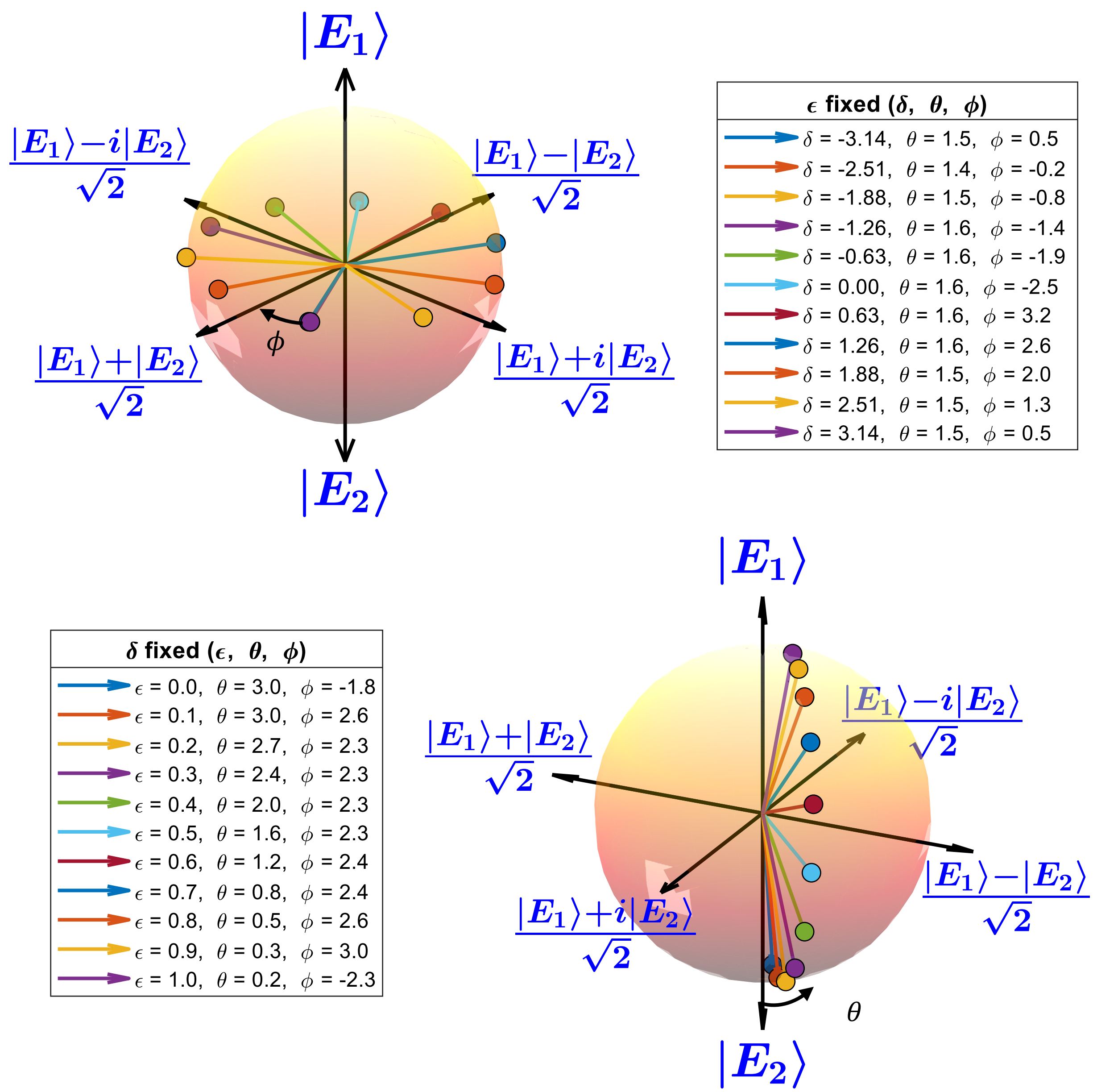

**Figure 4. Steady-State Bloch Sphere Representation:** Evolution of the system's state on the Bloch sphere as a function of the forcing parameters $\epsilon$ and $\delta$. **Left Panel**: For a fixed $\epsilon = 0.5$, varying $\delta$ from $-\pi$ to $\pi$ results in a complete revolution of the state vector in the same plane on the Bloch sphere. **Right Panel**: For a fixed $\delta = \pi/2$, varying $\epsilon$ from 0 to 1 causes the state to transition from eigenstate $|E_1\rangle$ to $|E_2\rangle$, tracing a path along the Bloch sphere and demonstrating the system's flexibility in state preparation.

For this analysis, we set $\epsilon = 0.5$, which provides a balanced contribution from both eigenmodes $|E_1\rangle$ and $|E_2\rangle$. By changing $\delta$, we achieve full rotational motion in the same plane, demonstrating that the relative phase $\delta$ plays a crucial role in defining the state trajectory while maintaining the amplitude ratio constant. By choosing different values of $\epsilon$ we can explore the full rotation of a plane that is more weighted to either $|E_1\rangle$ or $|E_2\rangle$.

Next, we varied $\epsilon$ while keeping $\delta$ fixed at $\pi/2$. This corresponds to the situation shown in the bottom panel of Fig. 4, where the system evolves from $|E_2\rangle$ and $|E_1\rangle$ as $\epsilon$ changes from 0 to 1. As $\epsilon$ increases, the system's state shifts from one eigenmode to the other, with $\alpha$ and $\beta$ continuously adjusting their relative magnitudes to reflect this transition.

In this case, the parameter $\epsilon$ governs the relative amplitude of the two driving forces. At $\epsilon = 0$, the system is completely in $|E_2\rangle$, and at $\epsilon = 1$, the system is fully in $|E_1\rangle$. As $\epsilon$ varies between these extremes, the state vector moves along a path from one eigenstate to the other, sweeping through an arc on Bloch sphere. This flexibility allows us to explore, for a given $\delta$, the full range of possible states by adjusting $\epsilon$, providing a powerful means of controlling the system's behavior.

By varying $\delta$ for a fixed $\epsilon$, we observe a complete revolution of the state vector within a plane of the Bloch sphere (left panel), demonstrating the effect of the relative phase between the two forces. When $\epsilon$ is varied while $\delta$ is kept constant at $\pi/2$, the system's state evolves between the eigenstates $|E_1\rangle$ and $|E_2\rangle$, effectively tracing a path along the surface of the Bloch sphere (bottom panel). This ability to reconstruct the system's state vector in a controlled manner provides insights into the underlying dynamics and allows for precise tuning of the system's behavior.

In the course of our measurements, we acknowledged potential sources of measurement uncertainty, such as instrumental precision, calibration errors, and environmental factors. To quantify uncertainty, we performed multiple repeated trials under the same conditions to assess the repeatability of the results. The system showed consistent behavior with minimal variation in the measured eigenfrequencies and amplitude/phase responses. The standard deviation of the repeated measurements was calculated and found to be within an acceptable range, once the system was properly calibrated and ensured that the system remained within the linear regime. A more detailed quantitative analysis of the Bloch angles, including the mean, standard deviation, and visualization of individual runs, is provided in the Supplementary Material under the subsection "Uncertainty and Repeatability of Bloch-Sphere Measurements.

## 4. Concluding Remarks

In this work, we experimentally demonstrated Bloch-state construction in a minimal linear mechanical platform consisting of two masses coupled by three springs. By operating the system at a fixed driving frequency between the symmetric and antisymmetric eigenfrequencies, the steady-state response was restricted to a two-mode manifold formed by the corresponding eigenmodes. Independent control over the relative amplitude and phase of the boundary forcing enabled systematic preparation of modal superpositions, whose complex coefficients were extracted from laser vibrometer measurements and mapped onto Bloch sphere coordinates. This procedure allowed direct reconstruction of the steady-state two-mode manifold and visualization of its geometric configuration on the Bloch sphere.. In contrast to previous systems, our setup ensures consistent repeatability and stability, making it a more reliable testbed for studying Bloch sphere trajectories.

Within this framework, sweeping the relative phase produced planar rotations on the Bloch sphere, while tuning the amplitude ratio continuously steered the system between eigenstate-dominated limits. The agreement between the measured modal coefficients, the reconstructed Bloch sphere trajectories, and the expected linear response confirms that the observed dynamics arise from controlled superposition within the two-mode subspace. Importantly, the present system isolates the geometric structure of superposition of states manipulation in a purely linear setting, without relying on nonlinear interactions, spectral mixing, or complex architectures.

These results establish a simple and experimentally transparent mechanical platform for programmable geometric state preparation. Beyond providing a benchmark for modal-superposition frameworks used in elastic-bit and pseudospin systems, the approach offers a controllable classical setting for exploring dynamics, phase-controlled state preparation, and quantum analogous Bloch state reconstruction strategies in coupled mechanical oscillators. Future work will focus on quantitatively analyzing the resonance curves, extracting damping values, and comparing theoretical Bloch trajectories with experimental results.

**Acknowledgements**

This work was partially supported by the Science and Technology Center New Frontiers of Sound (NewFoS) through the U.S. National Science Foundation (NSF) cooperative agreement #2242925. M.A.H also acknowledges the partial support from NSF Grant #2204382. P.A.D. and J.A.L. also acknowledge the partial support from NSF Grant #2204400.

**Author Contributions**

M.A.H., P.A.D., K.R., and J.A.L. conceived the idea of the research. A.N.F. and D.J.M. fabricated the samples, built the experimental setups, and performed the measurements. A.N.F. and D.J.M wrote the initial draft of the manuscript, with significant contributions from all authors. All authors analyzed the experimental data and participated in the scientific discussion.

**Competing interests**

The authors declare no competing interests.

**Data Availability**

The data that support the findings of this study are available from the corresponding author upon reasonable request.

**Nomenclature:**

| Symbol | Meaning |
|---|---|
| $F_1$ | Complex driving force applied at the right boundary |
| $F_2$ | Complex driving force applied at the left boundary |
| $\epsilon$ | State-ratio parameter governing relative magnitude of driving forces |
| $\delta$ | Relative phase between the two driving forces |
| $A_j$ | Complex steady-state response amplitude for mass $j$ |

| Symbol | Meaning |
|---|---|
| $\sigma_j$ | Absolute phases $\sigma_1$ and $\sigma_2$ for mass $j$ |
| $\alpha$ | Modal coefficient corresponding to the eigenmode $\|E_1\rangle$ |
| $\beta$ | Modal coefficient corresponding to the eigenmode $\|E_2\rangle$ |
| $\omega_D$ | Driving frequency |
| $\gamma_j$ | Phase of the driving force $F_j$ |
| $V_j$ | Voltage amplitude applied to channel $j$ of the function generator |
| $\theta$ | Polar angle in the Bloch sphere representing the relative modal amplitude |
| $\phi$ | Azimuthal angle in the Bloch sphere representing the phase difference between the eigenmodes |

## Supplementary Material

We consider a two-mass, three-spring system with damping, described by the following equations of motion (EOM):

$$M_1\ddot{u_1} = k_s\left(F_1 e^{i\omega_D t} - u_1\right) - k_s(u_1 - u_2) - \eta\dot{u_1}$$
$$M_2\ddot{u_2} = -k_s\left(u_2 - F_2 e^{i\omega_D t}\right) + k_s(u_1 - u_2) - \eta\dot{u_2} \quad \text{(S1)}$$

Here, $u_1$ and $u_2$ represent the displacements of the two masses, $F_1$ and $F_2$ are the driving forces applied at the two boundaries, $k_s$ is the spring constant, $\omega_D$ is the driving frequency, and $\eta$ is the damping coefficient. The system is weakly damped, and the goal is to linearize the equations to extract the system's response.

To simplify the analysis, we assume solutions of the form: $u_1 = A_1 e^{i\omega_D t}$ and $u_2 = A_2 e^{i\omega_D t}$ and applying in Eq. (S1), we derive the simultaneous equation for the amplitude $A_1$ and $A_2$ as,

$$\Rightarrow (-m\omega_D^2 + 2k_s + i\eta\omega_D)A_1 - k_s A_2 = k_s F_1$$
$$\Rightarrow -k_s A_1 + (-m\omega_D^2 + 2k_s + i\eta\omega_D)A_2 = k_s F_2 \quad \text{(S2)}$$

By taking the real part of the coefficient of the amplitudes from Eq. (S2), we can write the stiffness matrix of the solution:

$$[K] = \begin{bmatrix} -m\omega_D^2 + 2k_s & -k_s \\ -k_s & -m\omega_D^2 + 2k_s \end{bmatrix} \quad \text{(S3)}$$

The eigenmode frequencies of the system are determined by the stiffness matrix. To find the eigenmode frequencies, we solve the characteristic equation, where the determinant of the stiffness matrix is set to zero, $\det(K) = 0$. This yields the eigenmode frequencies as:

$$\omega_{01} = \pm\sqrt{\frac{k_s}{m}} \qquad \omega_{02} = \pm\sqrt{\frac{3k_s}{m}} \quad \text{(S4)}$$

From the solution of the eigenmode frequencies, we observe that the frequencies are influenced by the masses of the components and the linearized stiffness between them. By setting $k_s = 1$ and $m = 1$, we obtain the eigenmode frequencies as $\omega_{01} = 1$ and $\omega_{02} = \sqrt{3}$. Using the elimination and substitution method on the coupled amplitude equations from Eq. (S2), we derive the complex amplitude solutions $A_1$ and $A_2$.

$$A_1 = \frac{F_1 k_s(-m\omega_D^2 + 2k_s + i\eta\omega_D) + F_2 k_s^2}{(-m\omega_D^2 + 2k_s + i\eta\omega_D)^2 - k_s^2} \quad \text{(S5)}$$

$$A_2 = \frac{F_1 k_s^2 + F_2 k_s(-m\omega_D^2 + 2k_s + i\eta\omega_D)}{(-m\omega_D^2 + 2k_s + i\eta\omega_D)^2 - k_s^2}$$

Using the complex amplitude ($A_1$ and $A_2$) and the eigen states of $\left(E_1 = \frac{1}{\sqrt{2}}\begin{pmatrix}1\\1\end{pmatrix}\text{ and } E_2 = \frac{1}{\sqrt{2}}\begin{pmatrix}1\\-1\end{pmatrix}\right)$, we can express the displacement field of the masses through the superposition of the states' coefficients $\alpha$ and $\beta$:

$$\vec{U} = \begin{pmatrix}u_1\\u_2\end{pmatrix} = \begin{pmatrix}A_1\\A_2\end{pmatrix} e^{i\omega_D t} = \frac{1}{\sqrt{|\tilde{\alpha}|^2 + |\tilde{\beta}|^2}}\left(\tilde{\alpha}|E_1\rangle + \tilde{\beta}|E_2\rangle\right)e^{i\omega_D t}$$
$$= (\alpha|E_1\rangle + \beta|E_2\rangle)e^{i\omega_D t}$$
$$\Rightarrow \begin{pmatrix}A_1\\A_2\end{pmatrix} = \frac{1}{\sqrt{2}}\begin{pmatrix}\alpha+\beta\\\alpha-\beta\end{pmatrix} \tag{S6}$$

Substituting the complex amplitude ($A_1$ and $A_2$), we obtain the superposition of states coefficient $\alpha$ and $\beta$:

$$\alpha = \frac{A_1 + A_2}{\sqrt{2}} = \frac{k_L(F_1 + F_2)}{\sqrt{2}(-m\omega_D^2 + k_L + i\eta\omega_D)}$$
$$\beta = \frac{A_1 - A_2}{\sqrt{2}} = \frac{k_L(F_1 - F_2)}{\sqrt{2}(-m\omega_D^2 + 3k_L + i\eta\omega_D)} \tag{S7}$$

Using the two possible eigenmode vibration forms ($|E_1\rangle$ and $|E_2\rangle$), the displacement field vector $\vec{U}$ of the coupled masses, expressed in terms of the complex amplitude coefficients, can be written as:

$$\vec{U} = \alpha|E_1\rangle + \beta|E_2\rangle$$
$$\Rightarrow \vec{U} = r_\alpha e^{i\xi_\alpha}|E_1\rangle + r_\beta e^{i\xi_\beta}|E_2\rangle \tag{S8}$$

Here, $r_\alpha$ and $r_\beta$ represent the magnitudes of the complex amplitude coefficients for $\alpha$ and $\beta$, respectively. These states are expressed in polar coordinates as $r_\alpha = \cos\theta$ and $r_\beta = \sin\theta$, where $\theta$ is the polar angle, indicating the angle of the superposition states relative to the z-axis of the Bloch sphere, as illustrated in Fig. S1.

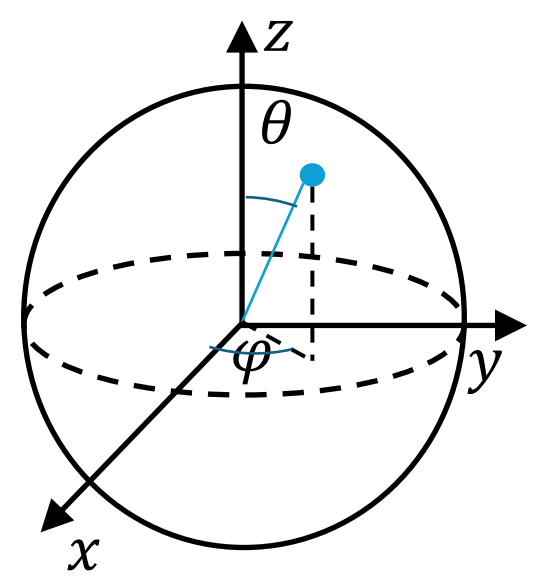

**Figure S1:** Three-dimensional illustration of the Bloch sphere, showcasing its relationship with the Cartesian coordinate system. This diagram highlights the sphere's axes and the defining angles—theta ($\theta$) and phi ($\phi$) representing an elastic bit's state

In Fig. S1, the z-axis of the Bloch sphere corresponds to the pure states $| E_1\rangle$ and $| E_2\rangle$, which are associated with the polar angle. By multiplying both sides of Eq. (S8) by the global phase factor $e^{-i\xi_\alpha}$, we can express the displacement field as $\vec{U}' = e^{-i\xi_\alpha}\vec{U}$. This allows us to represent the state space of the system accordingly.

$$\begin{aligned}\vec{U}' &= r_\alpha|E_1\rangle + r_\beta e^{i(\xi_\beta-\xi_\alpha)}|E_2\rangle \\ \vec{U}' &= \cos\theta\,|E_1\rangle + \sin\theta\, e^{i\phi}|E_2\rangle\end{aligned} \tag{S9}$$

Here, the phase difference between the complex amplitude coefficients is referred to as the azimuthal angle $\varphi = \xi_\alpha - \xi_\beta$, which signifies the rotation of the states in the x-y plane as depicted in Fig. S1. We limit the polar angle $\theta$ in the Bloch sphere to the range $0 \leq \theta \leq \pi$ because this range is enough to represent every possible pure qubit state without unnecessary repetition. A general qubit state can be written as $\vec{U} = \cos\left(\frac{\theta}{2}\right)|E_1\rangle + e^{i\phi}\sin\left(\frac{\theta}{2}\right)|E_2\rangle$, where $\theta$ determines the relative probabilities of measuring $|E_1\rangle$ and $|E_2\rangle$, while $\phi$ determines the relative phase between them. When $\theta = 0$, the state is $|E_1\rangle$, which corresponds to the north pole of the Bloch sphere, and when $\theta = \pi$, the state is $|E_2\rangle$ up to a global phase, corresponding to the south pole. If we allowed $\theta$ to go beyond $\pi$, for example up to $2\pi$, we would not get new physical states; instead, we would describe the same points on the sphere again using different angle values. Therefore, the restriction $0 \leq \theta \leq \pi$ is a coordinate convention that avoids double-counting states, not a physical limitation on the elastic-bit.

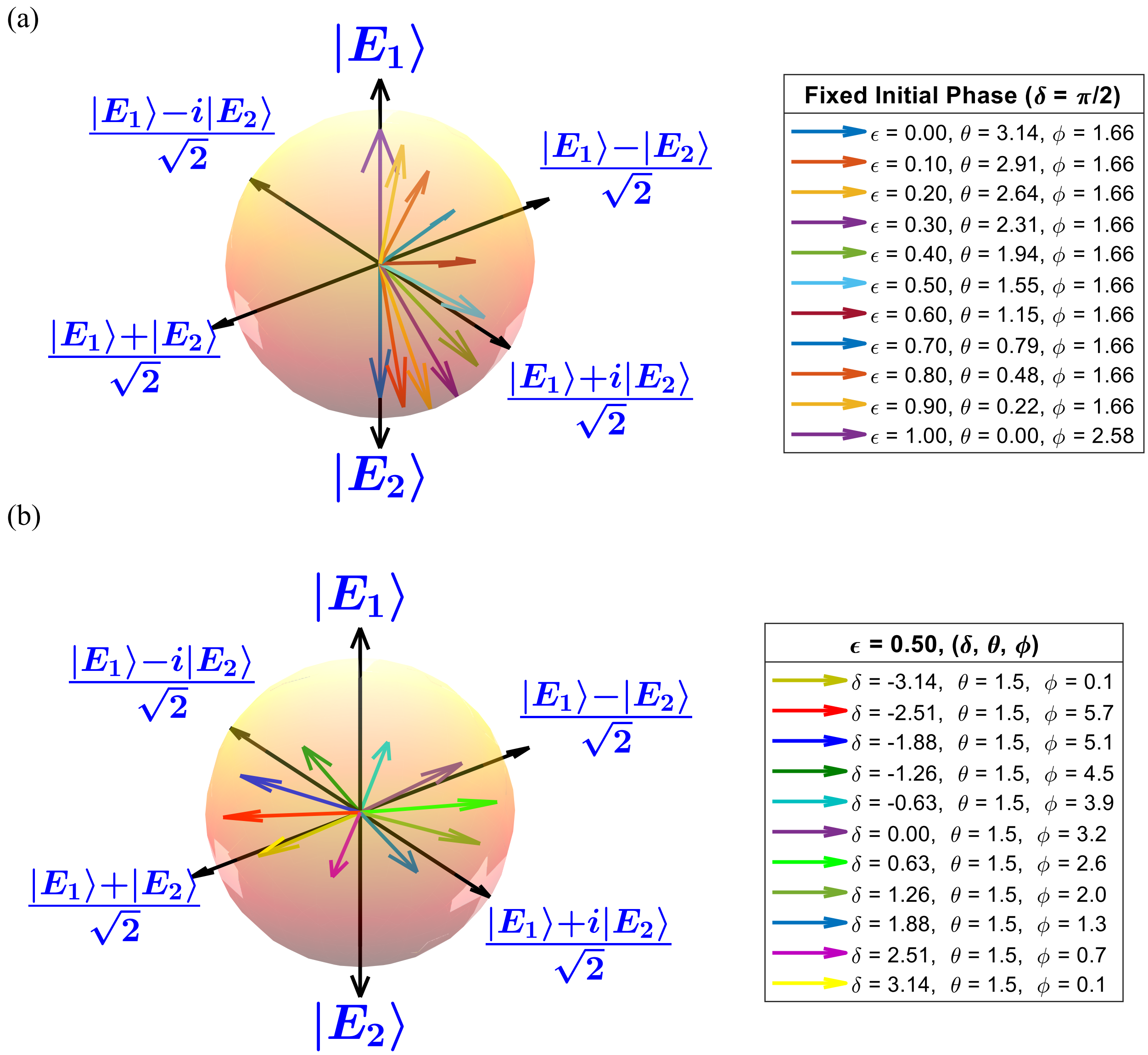

**Figure S2.** Illustration of the transformation of states of the two-level elastic bit system, represented geometrically on the Bloch sphere. Panel (a) shows the variation in the polar angle ($\theta$) with the ratio of $\epsilon$ while keeping the azimuthal angle ($\phi$) constant, illustrating the transition from pure states $|E_1\rangle(\theta = 0)$ to $|E_2\rangle(\theta = \pi)$ and various superpositions in between. The initial phase $\delta$ was fixed during this demonstration. Panel (b) focuses on the azimuthal angle ($\phi$) while keeping $\theta$ constant, by varying $\delta$ from $-\pi$ to $+\pi$ for a set value of $\epsilon = 0.5$. System parameters: $m = 1$, $k_s = 1$, $\eta = 0.03$, $\omega_D = \sqrt{2}$.

To investigate the behavior of the elastic bit on the Bloch sphere, we adjust the parameters $\epsilon$ and $\delta$ as described in Eq. (1). Figure S2 shows how these changes affect the polar angle ($\theta$) and azimuthal angle ($\phi$). We can visually trace the evolution of these angles on the Bloch sphere. This visualization demonstrates how varying $\epsilon$ and $\delta$ influences the superposition states of the elastic bit. Specifically, it reveals the range and nature of superposition states achievable through parameter manipulation. The Bloch sphere thus serves as a useful tool for understanding how different parameters alter the system's states.

To investigate scenarios where the azimuthal angle $\phi$ remains constant while the polar angle $\theta$ varies, we keep $\delta$ fixed and adjust the ratio of $\epsilon$. At $\theta = 0$, a pure state of $|E_1\rangle$ is achieved, and at $\theta = \pi$, a pure state of $|E_2\rangle$ is obtained (see Fig. S2(a)). As $\theta$ changes from 0 to $\pi$ [Fig. S2(a)], it signifies the generation of any superposition of states, incorporating both pure states $|E_1\rangle$ and $|E_2\rangle$.

We then explore another scenario where $\theta$ is fixed and $\phi$ varies, achieved by different parameter adjustments. As demonstrated in Fig. S2(b), by setting $\epsilon = 0.5$ and varying $\delta$ from $-\pi$ to $+\pi$, we observe a constant $\theta$ while $\phi$ completes a full $2\pi$ rotation.

**Linearity of the Two-Mass System**

To support the claim of linear behavior in the two-mass, three-spring system, we performed detailed sweeps of the boundary driving conditions across the full operating range used for the Bloch-sphere reconstruction. The system was analyzed under two variations of the forcing parameters: the state-ratio parameter $\epsilon$, which was varied from 0 to 1, and the relative driving phase $\delta$, which was varied from $-180°$ to $180°$. For each case, the velocity responses of Mass 1 and Mass 2 were recorded, and the phase difference between the masses was calculated. The driving frequency was fixed at 99 Hz, between the in-phase and out-of-phase eigenfrequencies, and each measurement included a brief stabilization period of 10 seconds to ensure steady-state conditions. For the $\epsilon$sweep, the amplitudes of the boundary excitations were scaled according to $\epsilon$and the complementary amplitude ratio, and the maximum voltage applied to each boundary was limited to 1 $V_{\mathrm{pp}}$ to remain within the linear regime. For the $\delta$ sweep, the relative phase between the two boundary excitations was varied while keeping the amplitude ratio fixed at $\epsilon = 0.5$. The complex amplitude of each mass was extracted using FFT analysis, and the inter-mass phase difference was computed relative to the synchronized driver reference.

The results show that the amplitudes of Mass 1 and Mass 2 scale predictably with $\epsilon$, while the phase difference remains nearly constant across the full range, demonstrating stable linear behavior.

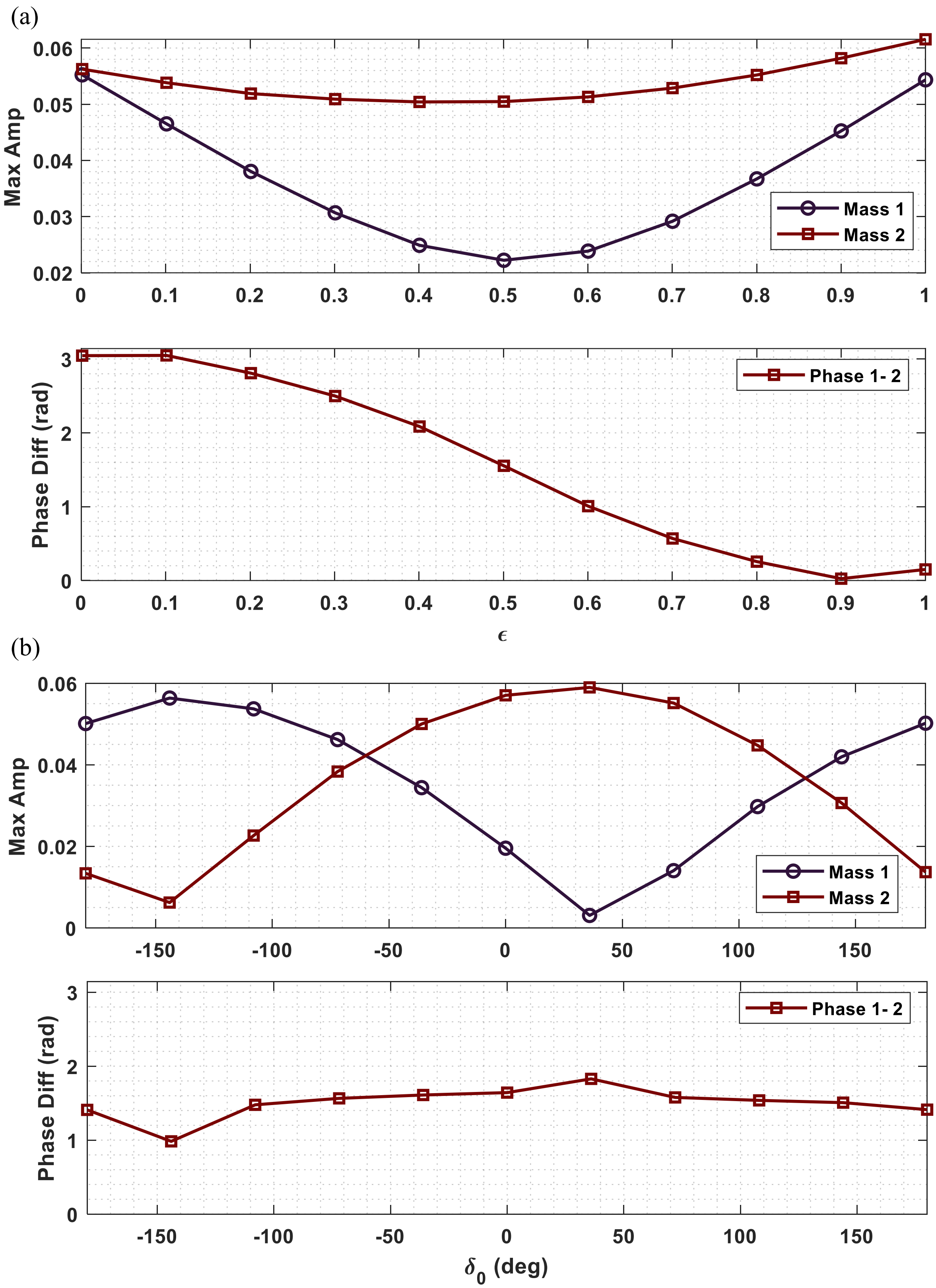
(a)
Max Amp
Mass 1
Mass 2
Phase Diff (rad)
Phase 1- 2
ϵ
(b)
Max Amp
Mass 1
Mass 2
Phase Diff (rad)
Phase 1- 2
δ0 (deg)

**Figure S3:** (a) Amplitude and phase difference of Mass 1 and Mass 2 as a function of the state-ratio parameter $\epsilon$. (b) Amplitude and phase difference of Mass 1 and Mass 2 as a function of the relative driving phase $\delta$. These plots demonstrate the predictable scaling of amplitudes and the stability of phase differences across the operating range, confirming the linear behavior of the system.

Similarly, varying the relative phase $\delta$ produces consistent amplitude oscillations and stable phase differences between the masses, confirming linearity across the operating range used for the Bloch-sphere experiments. These findings provide quantitative evidence that the system operates in a linear regime throughout the experimental conditions, supporting the validity of the steady-state geometric state reconstruction presented in the main manuscript. The results are summarized in Figure S3, where panel (a) corresponds to the $\epsilon$ sweep and panel (b) correspond to the $\delta$ sweep.

**Uncertainty and Repeatability of Bloch-Sphere Measurements**

To quantify the repeatability of the Bloch-sphere reconstruction, we performed 10 independent runs under the same driving condition ($\epsilon = 0.5$, $\delta = 0.8 * \pi$). For each run, the Bloch angles $\theta$ and $\phi$ were extracted from the velocity data of the two masses and converted into Cartesian coordinates for visualization on the Bloch sphere.

The mean values across the 10 runs were used to represent the nominal state, while the standard deviations of the measurements were calculated to assess variability. For the 10 runs considered, the average repeatability error was found to be $1.2\%$ for $\theta$ and $2.7\%$ for $\phi$. These results demonstrate that the reconstructed states are highly consistent across repeated measurements.

The variability in $\theta$ and $\phi$ is shown in Figure S4(a) as histograms overlaid with Gaussian fits, providing a clear quantitative representation of the spread of the measurements. The Bloch-sphere representation in Figure S4(b) shows the mean vector along with the individual vectors from all ten runs. Each individual run is represented by a vector from the origin to the tip, while the mean vector is highlighted, giving an intuitive visualization of the uncertainty and repeatability of the reconstructed Bloch-sphere states.

Together, these analyses provide quantitative evidence for both the uncertainty and repeatability of the measurements, confirming that the system produces consistent and reproducible Bloch-sphere trajectories under the specified driving conditions.

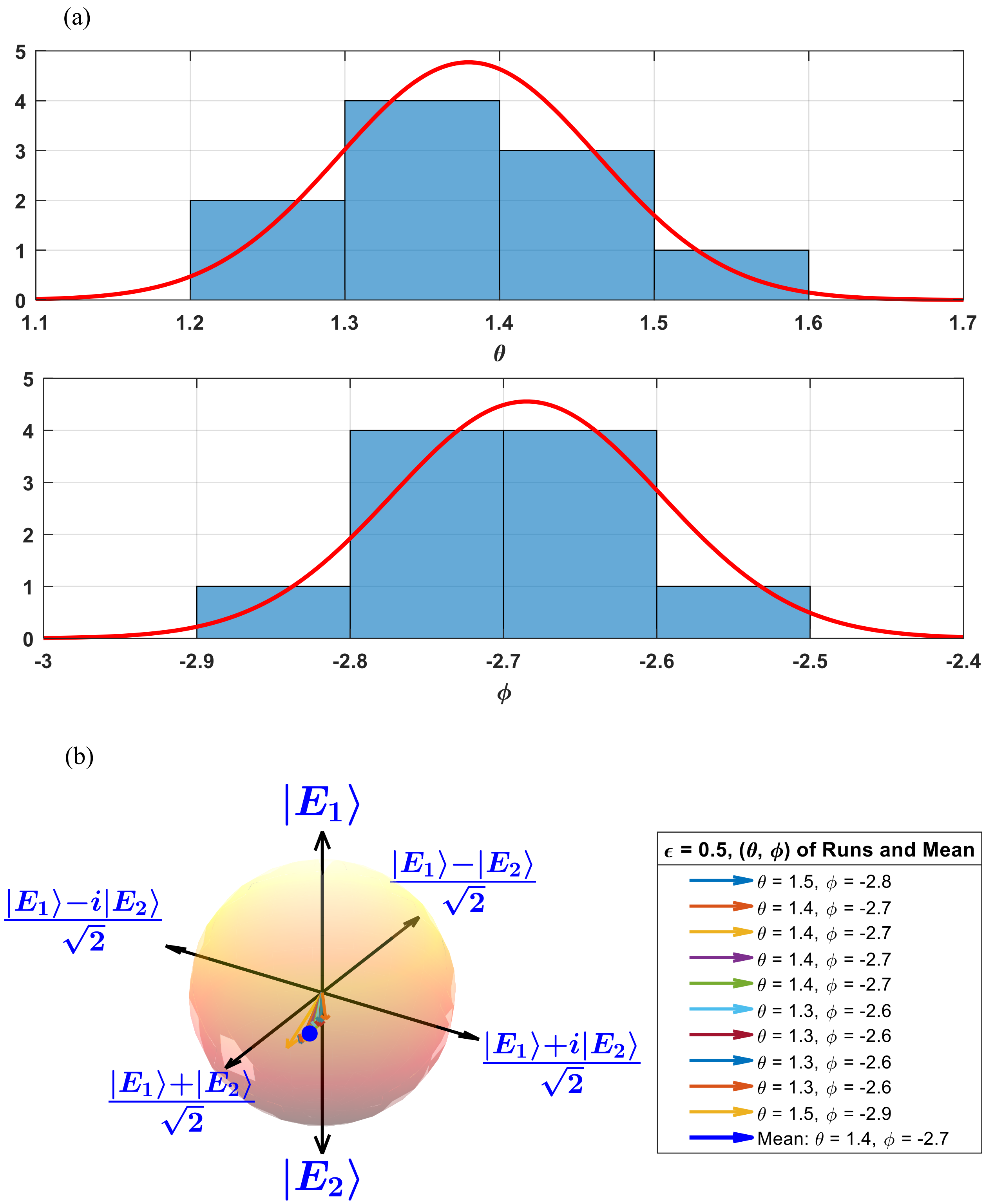

**Figure S4:** Quantitative assessment of uncertainty and repeatability for the Bloch-sphere reconstruction under the driving condition $\epsilon = 0.5$, $\delta = 0.8 * \pi$. (a) Histograms of $\theta$ and $\phi$ values across ten independent runs with Gaussian fits, illustrating the spread of the measurements and standard deviations. (b) Bloch-

sphere representation showing the mean vector along with all individual vectors from the ten runs, visualizing the variability and confirming the repeatability of the reconstructed states.